\documentclass[12pt]{article}
\usepackage{epsfig}

\parskip=\medskipamount	
\renewcommand{\thesection}{\arabic{section}}

\catcode`@=11

\@addtoreset{equation}{section}
\@addtoreset{equation}{subsection}
\def\theequation{\ifnum\value{section}=0 \arabic{equation}\ignorespaces
\else \ifnum\value{section}=-1 A.\arabic{equation}\ignorespaces
\else \ifnum\value{subsection}=0 \thesection.\arabic{equation}\ignorespaces
\else \thesection.\arabic{subsection}.\arabic{equation}\ignorespaces
                             \fi
                        \fi
                   \fi}

{\catcode`\'=\active \def'{{}^\bgroup\prim@s}}

\catcode`@=12

\newcommand{\bq}{\begin{equation}}
\newcommand{\be}{\begin{equation}} 
\newcommand{\fq}{\end{equation}}
\newcommand{\ee}{\end{equation}}
\newcommand{\bea}{\begin{eqnarray}}
\newcommand{\eea}{\end{eqnarray}}

\def\bop#1{\setbox0=\hbox{$#1M$}\mkern1.5mu
	\vbox{\hrule height0pt depth.04\ht0
	\hbox{\vrule width.04\ht0 height.9\ht0 \kern.9\ht0
	\vrule width.04\ht0}\hrule height.04\ht0}\mkern1.5mu}
   
\begin{document} 

\thispagestyle{empty}

\begin{flushright}
\begin{tabular}{l}
CTP-MIT-3103\\ 
 
\end{tabular}
\end{flushright}

\vskip.3in
\begin{center}
{\Large\bf Aspects of Gauge Theory on Commutative \\ and
	 Non-commutative Tori}
\vskip.3in

\vskip .2in
{\bf Zachary Guralnik and  Jan Troost}
\\[5mm]
{\em Center for Theoretical Physics \\
Massachusetts Institute of Technology\\
Cambridge MA, 02139}\\
{Email: zack@mitlns.mit.edu,troost@mit.edu}

\vskip.5in minus.2in

{\bf Abstract}

\end{center}

We study aspects of gauge theory on tori which
are a consequences of Morita equivalence.  
In particular we study the behavior of gauge theory on non-commutative 
tori for arbitrarily close rational values of $\Theta$. 
For such values of $\Theta$,  there are Morita equivalent 
descriptions in terms of Yang-Mills theories on commutative tori
with very different magnetic fluxes and rank.
In order for the correlators of open Wilson lines to 
depend smoothly on $\Theta$, the correlators of closed 
Wilson lines in the commutative Yang-Mills theory must satisfy  
strong constraints. If exactly satisfied,  these constraints 
give relations between small and large $N$ gauge theories.   
We verify that these constraints are obeyed at leading order
in the $1/N$ expansion of pure 2-d QCD and 
of strongly coupled ${\cal N} = 4$ super Yang-Mills theory.

\setcounter{page}{0}  
\newpage 
\setcounter{footnote}{0}


\section{Introduction}

The subject of gauge theory on non-commutative spaces has received much
attention recently for a variety of reasons.  In string theory, interest
in the subject was kindled by \cite{CDS} where it arose in the context
of toroidal compactification of M(atrix) theory corresponding to eleven
dimensional supergravity
backgrounds with a constant three-form field.
In \cite{SW},  it was shown that non-commutative gauge theories arise 
naturally in IIA/B string theory as a certain decoupling limit of D-branes
in NS-NS two-form backgrounds.
In this paper we shall be interested in the behavior of gauge theories
on non-commutative tori.  In particular we shall study the question of
how smoothly the non-commutative theory depends on the non-commutativity
parameter $\Theta$ on a spacelike $T^2$.  Our motivation for addressing
this question is that continuity in $\Theta$ has strong 
consequences for the behavior of commutative gauge theories,  due to the
Morita equivalence \cite{Schwarz} 
between commutative and non-commutative spaces for 
rational $\Theta$.
  
The question of smoothness in $\Theta$ was addressed in part in 
\cite{EPR},  following the work of
\cite{AI}.  These authors noticed different high energy behavior 
for rational and irrational $\Theta$.
As one increases the energy,  there is a sequence of T-dual
(or Morita equivalent) descriptions such that the theory has a 
quasi-local description.  In the rational case,  this sequence terminates 
above some finite energy,  whereas in the irrational case the sequence never 
terminates.  Since any irrational number is arbitrarily 
close to a rational one,  
the theory does not depend smoothly on $\Theta$ from this point of view.  
On the other hand,  if one varies $\Theta$ at a fixed energy,
then the behavior is essentially smooth.  Only at certain isolated values of 
$\Theta$ does the quasi-local T-dual description change. Note that these
are not really phase transition points,  and so do not represent any
non-smooth behavior of the physics.   

In fact, whether or not things depend smoothly on $\Theta$ seems to 
depend on the question one asks.  For instance, the moduli space of 
flat connections is independent of $\Theta$ \cite{KS2} 
(and therefore depends smoothly on $\Theta$).   However
the  periodicity of non-commutative gauge fields for instance does not 
depend smoothly on $\Theta$. 
A natural question is whether correlation functions open Wilson lines
with fixed transverse momenta depend continuously on $\Theta$.

One way to address the question of continuity is to consider 
arbitarily close rational values of $\Theta$.  For rational 
$\Theta$ there is a 
Morita equivalent description in terms of a $U(N)$ Yang-Mills
theory with magnetic flux $m$ on a commutative torus 
\cite{CDS, SW, PS, B}.  The rank $N$ and flux $m$ of the commutative 
description differ drastically for arbitarily close rational values of 
$\Theta$.  If for instance one considers the $U(1)$ gauge theory on
a non-commutative $T^2$ with $\Theta = -c/N$ for integer $c$ and $N$,  
then the dual commutative theory has rank $N$ and flux $m$ which are 
solutions of $a N - c m = 1$, where $a$ is integer.   
Thus the commutative dual for $\Theta = -1/2$ is 
a $U(2)$ gauge theory,  whereas the commutative dual for $\Theta = 
-1000/2001$ is an $U(2001)$ gauge theory.  It follows that for open 
Wilson line correlators to depend smoothly
on $\Theta$ -- up to some possible isolated phase transition 
points in $\Theta$ --,  there must 
be relations between commutative gauge theories of very different 
rank.  For this reason,  smooth dependence of correlators on 
$\Theta$ is a very non-trivial phenomenon.  Yet continuity seems likely,
at least in cases in which the non-commutative Yang-Mills theory 
is obtained from string theory,  where the NS-NS two form is a 
continuous background.  Comparing the small and large $N$ gauge 
theories is beyond the scope of this paper.  However we will
be able to check some of the properties implied by smooth $\Theta$ 
dependence in the large $N$ expansion of commutative gauge 
theories.  

The T-duality which relates the non-commutative theory
to the commutative one maps open Wilson lines of the non-commutative
theory to closed Wilson lines wrapping the torus of the 
commutative theory \cite{AMNS,S}.  Thus the condition 
that the open Wilson line correlators depend
smoothly on $\Theta$ can be translated into a condition satisfied
by correlators of wrapped Wilson loops in the commutative theory.
The latter can only depend on certain 
combinations of $m$, $N$, the radius  $R$, the coupling
 $g^2$, and the winding number 
and transverse momenta of the wrapped Wilson loops. 
(In asymptotically free theories,  $g^2$ is replaced by the QCD scale.)
We  find that the Wilson loop correlators have the required
dependence on these variables at leading order in the $1/N$ expansion
of the strongly coupled four dimensional ${\cal N} =4$ super Yang-Mills 
theory.  In pure two-dimensional Yang-Mills,  we shall also find the correct 
behavior at leading order in the $1/N$ expansion.  
In both cases, the smooth behavior depends crucially on the fact that
the magnetic flux behaves as a $B$ field modulus in a string description
of the Yang-Mills theory.

The organization of this paper is as follows.
In section \ref{review} we review the definition of gauge theory on a non-commutative
torus.  In section \ref{morita}, we give a simple construction of the Morita equivalence
between a $U(1)$ gauge theory compactified on a non-commutative $T^2$ with 
rational $\Theta$ and a $U(N)$ Yang-Mills theory on a commutative torus with 
magnetic flux.  Morita equivalence was explicitly shown in generality
in \cite{PS}.  The non-smooth dependence of the gauge field periodicity on
$\Theta$ follows simply from this map. In section \ref{observables} we  construct the map
between open Wilson lines \cite{AMNS,S} (including possible local 
operator insertions) in
the non-commutative theory and wrapped Wilson lines in the commutative dual. 
In section \ref{correlators} we discuss the correlation functions of wrapped Wilson loops
in large N strongly coupled large ${\cal N} =4$ super Yang-Mills theory using
the AdS-CFT correspondence, and show that at least to leading order in 
$1/N$ these have the behavior predicted by continuity in $\Theta$.
We also discuss confining Yang-Mills theory 
in the $1/N$ expansion,  and find behavior consistent with continuity in $\Theta$. In section \ref{cd} we discuss other possible tests of smooth behavior,
and give our conclusions.  

\section{Review of Non-commutative Gauge Theories}
\label{review}

Gauge theory on a non-commutative space can be constructed
essentially by replacing regular products of functions with
Moyal star products.  The star product is defined as follows:
\be
f(x)*g(x) = e^{i \frac{\theta^{\mu\nu}}{2}
\frac{\partial}{\partial x^{\mu}}\frac{\partial}{\partial y^{\nu}}}
	f(x)g(y)|_{y\rightarrow x}.
\ee
The action of the non-commutative pure $U(p)$ Yang-Mills theory is
\be
S=\frac{1}{g^2} \int d^Dx 
Tr({\cal F}_{\mu\nu}(x) + \Phi_{\mu\nu}) * ({\cal F}_{\mu\nu}(x) + \Phi_{\mu\nu})
\ee
where 
\be
{\cal F}_{\mu\nu}(x) = \partial_{\mu}{\cal A}_{\nu} - 
\partial_{\nu}{\cal A}_{\mu} - 
i ({\cal A}_{\mu} * {\cal A}_{\nu} - 
	{\cal A}_{\nu}* {\cal A}_{\mu})
\ee
and $\Phi$ is a background U(1) term whose significance will become apparent
shortly. 
We shall be interested in the case in which two of the spatial 
directions $x^1$ and $x^2$ are non-commutative and compactified on a torus, 
i.e.
\be
x^1 * x^2 -x^2 * x^1 = i\theta
\ee
with $x^1 \equiv x^1 + 2\pi R$ and $x^2 \equiv x^2 + 2\pi R$.

The existence and renormalizability of these theories is still an open
question,  which has been addressed partially \cite{F}.
Certain non-commutative $U(N)$ gauge theories  
arise in a  decoupling limit \cite{SW} of D-branes in type IIA/B
string theory. At least in these cases the theory 
should exist microscopically rather than just as an effective field theory.
An alternative way to try to define the quantum theory uses
a lattice regularization \cite{AMNS}.

When  compactified on a non-commutative $T^2$,
these theories exhibit an $SO(2,2,Z)$ Morita equivalence which is 
inherited from string theory T-duality (see \cite{CDS, SW, mora}).  
Morita equivalence has also been demonstrated explicitly without 
recourse to either
string theory or supersymmetry \cite{PS,B, AMNS,S}.  
This equivalence exists at the classical level.
The duality group has an $SL(2,Z)$ subgroup which acts as follows: 
\bea
\pmatrix{ m \cr N } &\rightarrow&
\pmatrix{a & b \cr c & d} \pmatrix{m \cr N} \\
\Theta &\rightarrow& 
\frac{c + d\Theta}{a + b\Theta} \\
R^2 &\rightarrow& R^2(a+ b \Theta)^2 \\
g^2 &\rightarrow& g^2(a + b\Theta) \\
\Phi &\rightarrow& (a+b\Theta)^2 \Phi - b(a + b\Theta)
\eea
where $\Theta \equiv \theta/ (2\pi R^2)$ and $2\pi R$ is the circumference 
of the torus,  which for simplicity we take to be square. The magnetic
flux is denoted by $m$ and $N$ is the rank of the gauge group.
The background field $\Phi$ is equal to the NS-NS $B$-field in the $\alpha'
\rightarrow 0$  limit when $\Theta=0$.
If one starts with a non-commutative $U(1)$ gauge theory with 
rational $\Theta = -c/d$  (with $gcd(c,d)=1$) and vanishing first Chern class,
then there is an $SL(2,Z)$ transformation of the form 
\be
T = \pmatrix{a & b \cr c & d}
\ee
which gives rise to a standard gauge theory
with $\Theta =0$. 
Thus one could take the view that the non-commutative theory exists at
least for rational $\Theta$,  and then try to define the theory at irrational 
$\Theta$ by approaching it with
an infinite sequence of rational numbers.  This can only make sense
if observables of the theory are arbitrarily close for arbitrarily close
rational $\Theta$.  However as pointed out in \cite{CDS}, the continuity of 
observables in $\Theta$ should not be taken for granted.

\section{Morita equivalence of NC Yang-Mills theory}
\label{morita}

In this section we explicitly demonstrate 
Morita equivalence for $U(1)$ theories on a torus with rational $\Theta$. 
The explicit map has been constructed in 
\cite{PS,B, AMNS, S} \footnote{This map was already  
hinted at in \cite{zmap},  although the context of that paper was 
different.}.  
We review this map and illustrate some of the issues 
related to continuity in $\Theta$.  
  
We will consider the non-commutative
$U(1)$ Yang-Mills theory on $T^2 \times R^n$,  with 
$\Theta_{12} = -c/N$.
We take the first Chern class and background term
to vanish.  
For simplicity we 
consider a square torus, with sides of length $2\pi R'$.  
There is then an $SL(2,Z)$ T-duality
\be
T = \pmatrix{a & m \cr c & N}
\ee
which gives a commutative $U(N)$ gauge theory. The
gauge coupling and radius of the torus are mapped as follows:
\begin{eqnarray}
{g'}^2 = g^2 N \\
R' = N R,
\end{eqnarray}
where the unprimed quantities are in the commutative theory.

\subsection{Explicit Form of Morita equivalence}

We chose the first Chern class and background term to vanish in the 
non-commutative $U(1)$ gauge theory.
In this case, the dual $U(N)$ commutative theory has first Chern class
$\int_{T^2} Tr F= 2 \pi m$ and a
background field $\Phi$ such that  $\int_{T^2} Tr(F + \Phi) = 0$.
The action is
\begin{equation}
S = \frac{1}{g^2} \int tr (F - \frac{m}{2\pi N R^2} )^2.
\end{equation}
To demonstrate Morita equivalence explicitly we start with
the pure $U(N)$ gauge theory on a commutative
$T^2$. 
For a $U(N)$ bundle the first Chern class is equal to the 't Hooft 
magnetic flux \cite{tH} in the $SU(N)$ component  of the gauge group,
modulo $N$ (see e.g. \cite{GR}).
We can redefine the $U(1)$ part of
the gauge field to remove the background term,  giving a 
$U(1) \times SU(N)/Z_N$ bundle with 't Hooft flux $m$ and vanishing
first Chern class.  This redefinition is a matter of convenience
in writing down the explicit form of the Morita equivalence, as  
it will allow us
to do a Fourier transform of the gauge potential on a multiple cover
of the torus.    

The 't Hooft magnetic flux corresponds to twisted boundary conditions
on the torus,  characterized by a pair of transition functions.
For a 't Hooft flux $m$,  we may choose
the boundary conditions:
\bea
A(x^1+2\pi R,x^2) = P^{-m} A(x^1,x^2) P^{m} \\
A(x^1, x^2 + 2\pi R) = Q A(x^1,x^2) Q^{-1}
\eea
where  
\be
Q =
\pmatrix{ 1 & & & \cr
& e^{\frac{2\pi i}{N}} & & \cr
& & \ddots & \cr
& & & e^{\frac{2\pi i (N-1)}{N}}}
\ee
and
\be
P = 
\pmatrix{
 0 & 1 & & \cr
 & 0  & 1 & \cr
 &   &   & \ddots \cr
 1 &   &   &}  
\ee
$Q$ and $P$ satisfy 
\begin{equation}
PQ = QP e^{2\pi i\over N}.
\end{equation}
We restricted ourselves to the case in which $N$ and $m$
are relatively prime:
\begin{equation}
a N - c m = 1
\label{relprime}
\end{equation}
for some integers $a$ and $c$.  
Since $(P^m)^N = Q^N = 1$,  the gauge field
is strictly periodic on an $N^2$ cover of the torus,  with periods
$2\pi NR$.  Because the greatest common divisor of $N$ and $m$ is $1$,
there is no smaller multiple cover on which the gauge fields are periodic.

A general form for a gauge field satisfying the above boundary conditions 
is given by 
\begin{equation}
A^{\mu} = 
\sum_{\vec r} a^{\mu}_{\vec r} Q^{-c r_1}
P^{r_2} \exp \left(-i\pi {c\over N}r_1r_2 \right)
\exp \left( - i \frac{ {\vec r} \cdot {\vec x} } {NR}\right).
\label{genform}
\end{equation}
The phase factor $\exp{-i\pi {c\over N}r_1r_2}$ can be absorbed
in the definition of $a^{\mu}_r$,  but is included for convenience as it
will lead to an easily recognizable form for the dual action.  
The field strength 
$F_{\mu\nu} = \partial_{\mu}A_{\nu} 
- \partial_{\nu}A_{\mu} -i[A_{\mu},A_{\nu}]$, may be written as 
\begin{equation}
F^{\mu\nu} = \sum_{\vec r}  
f^{\mu\nu}_{\vec r} Q^{-c r_1}
P^{r_2} \exp \left(- i\pi {c\over N}r_1r_2\right)
\exp \left( -i {\vec r \cdot \vec x \over NR}\right),
\end{equation}
where
\bea
f^{\mu\nu}_r =  -{i\over NR} (r^{\mu}a^{\nu}_r - r^{\nu}a^{\mu}_r) 
\ \ \ \ \ \ \ \ \ \ \ \ \ \ \ \ \ \ \ \ \ \ \ \ \ \ \ \ \ \ \ \ \ \ \ \ \ \ \ 
\ \ \ \ \ \ \ \ \ \ \ \ \ \ \ \ \ \ 
\nonumber \\
	- i \sum_{r^{\prime}} a^{\mu}_{r-r^{\prime}} a^{\nu}_{r^{\prime}}
\left[ 
\exp \left( i\pi{-c\over N}
(r_2r^{\prime}_1 - r_1r^{\prime}_2) \right)
-\exp \left(i\pi{c\over N}(r_2r^{\prime}_1 - r_1r^{\prime}_2)\right) 
\right]
\eea

In terms of the modes $a^{\mu}_r$ and $f^{\mu\nu}_r$,  one can define 
a dual $U(1)$ gauge field which is periodic on a torus with radius
$R^{\prime} = NR$;
\begin{equation}
{\cal A} = \sum_r a_r \exp \left( -i \frac{{\vec r} \cdot {\vec x}}
{R^{\prime}} \right)
\end{equation}
and a field strength,
\begin{equation}
 {\cal F} =  \sum_r 
	f_r  \exp \left( - i \frac{{\vec r} \cdot {\vec x}} 
{R^{\prime}}\right).
\end{equation}
It is then straightforward to show that 
\begin{equation}
 {\cal F}_{\mu\nu}  = 
	\partial_{\mu}{\cal A}_{\nu} - \partial_{\nu}{\cal A}_{\mu} 
	-i ({\cal A}_{\mu} * {\cal A}_{\nu} - 
	{\cal A}_{\nu} * {\cal A}_{\mu})
\end{equation}
provided that 
\be
\theta = -\frac{c}{N} 2\pi {R^{\prime}}^2.
\ee
In terms of the dual variables,  the action is
\begin{equation}
S = \frac{1}{{g^{\prime}}^2 } \int_{{T^2}^{\prime} \times R^n} {\cal F}*{\cal F}.
\end{equation}
where the torus has sides of length $2\pi R^{\prime}$ and 
\begin{equation}
{g^{\prime}}^2 = g^2 N. 
\end{equation}
Note that the flux of the dual non-commutative $U(1)$ field strength vanishes,
and that there is no background field in the action.

One consequence of this map is that the non-commutative
gauge field ${\cal A}$ is a periodic variable,  with period
\be
{\cal A} \sim {\cal A} + \frac{N}{R^{\prime}}.
\label{period}
\ee
This follows from the periodicity of the commutative gauge field
$A \sim A +1/R$ and the Morita map.  At fixed $R'$, this period 
does not vary continuously with $\Theta$ since $N$ fluctuates
wildly between arbitarily close rational values of  $\Theta = -c/N$.

\section{Mapping observables}
\label{observables}

We wish to identify quantities in the non-commutative theory  
which vary smoothly as a function of
$\Theta$, keeping ${g'}^2$ and $R^{\prime}$ fixed.  
Via Morita equivalence,  the existence
of such quantities requires relations between small and large $N$ gauge 
theories.

\subsection{Wilson loop observables}

Among the quantities one might guess vary smoothly with $\theta$ are 
correlation functions of open Wilson lines (see \cite{IS, Rey1, GHI}) at fixed 
transverse momenta. 
In the non-commutative theory, open Wilson lines are gauge invariant
provided they have a certain transverse momentum.
For simplicity we will consider the gauge invariant operators built 
from straight open Wilson lines, given by
\begin{equation}
{\hat O}_p = \int d^2x W(x^i \rightarrow x^i +p_j\theta^{ij}) *  
\exp(i{\vec p} \cdot {\vec x})
\end{equation}
where $W(x \rightarrow y)$ is an open Wilson line, path ordered with 
respect to the non-commutative star product,  and stretching between 
the points $x$ and $y$ in the non-commutative plane.
Because of the non-commutativity,  the term $\exp(i{\vec p} \cdot {\vec x})$
is a translation operator in the direction transverse to $\vec p$,  which
for gauge invariance must relate the two endpoints of the open Wilson line
(see \cite{GHI} for a more detailed discussion).  
If one writes the path between the endpoints as $x(s) = x + \Delta s$,  
where $s$ is a parameter 
on the interval $[0,1]$,  then 
\bea
W &=& P_{\ast} \exp\left(i\int {\cal A}\right)  \\
&=& 1 +i \Delta \int_0^1 ds {\cal A}(x + s\Delta) - 
\Delta^2 \int_0^1 ds \int_0^s d \tilde s {\cal A}(x+ s\Delta) * 
{\cal A}(x+ \tilde s \Delta)  + \cdots, \nonumber
\eea
where we have path ordered from right to left. For concreteness
we consider Wilson lines stretching in the $2$ direction. 
Upon compactifying on a torus with a non-commutative parameter 
$\Theta = -\frac{c}{N} 2 \pi {R'}^2 $,  the gauge invariant open Wilson lines are 
\be
\hat O_{{n'}_1, {e'}_2} = 
\int d^2x Tr W(x^2,x^2 - {n'}_1\frac{c}{N} 2\pi R^{\prime} + 2\pi R' e_2') 
*  \exp(i \frac{ n_1' x^1}{R'}) 
\label{non-comwilson}
\ee
where $n_1'$ and $e_2'$ are integers. The operator  
(\ref{non-comwilson}) can be regarded as a creation operator for a state with 
momentum $n_1'$ and electric flux $e_2'$. Note that these operators are
invariant under ${\cal A} \rightarrow {\cal A} + N/R'$.   

The creation operator for states with 
't Hooft electric flux in the 
commutative Yang-Mills theory theory are Wilson lines wrapping the torus.
In the presence of magnetic flux,  these involve 
the transition functions $U_i$ \cite{VB} associated to the twisted boundary
conditions on the torus. 
For instance,  a Wilson line beginning at the point $(x^1, x^2)$ and wrapping 
the torus $e_2$ times in the $2$ direction is given by
\begin{equation}
\hat O_{e_2}(x^1) = 
Tr P \exp\left(i\oint_{x^2}^{x^2+ 2 \pi R e_2} 
A_2  dx^2\right) {U_2}^{-e_2}.
\label{comwils}
\end{equation}
We  consider
the Wilson loop in the fundamental representation,  so that $e_2$ is
the amount of electric flux created by $W$.
Roughly speaking,  the term $U^{-e_2}_2$ is dual to the factor
$\exp(i{\vec p} \cdot {\vec x})$ corresponding to the fractional momentum
in the expression for the open Wilson
line of the non-commutative theory.
The transition functions in this expression are necessary to ensure 
invariance of the Wilson loop under ``small'' gauge transformations
i.e. those which are generated by the Gauss law operator $D_i F_{0i}$.
The small gauge transformations $V_s$ are those which satisfy the same 
twisted boundary conditions as the field strength:
\begin{equation}
V_s(x^1,x^2 + L^2)  = 
U_2(x^1,x^2)V_s(x^1,x^2)U_2^{\dagger}(x^1,x^2).
\end{equation}
One can check that (\ref{comwils}) is independent of 
$x^2$, and that for a U(N) bundle, 
the Wilson loop (\ref{comwils}) is periodic in $x^1$ in spite of the twisted boundary 
conditions on the gauge field.

\subsection{Mapping Wilson loops}

One can  guess  how non-commutative open Wilson lines map under 
Morita equivalence, since, at least in the maximally supersymmetric case, this 
equivalence is a manifestation of T-duality 
in string theory.
Nevertheless one can explicitly construct this map in the 
Yang-Mills theory \cite{AMNS, S}, making no use of supersymmetry.
We discuss this map below.

Before writing the explicit transformation,  let us consider general 
properties which give physical intuition into the nature of the 
Morita map.
Consider the $Z_N \times Z_N$ set of large gauge 
transformations in the $SU(N)$ component of the commutative theory. 
These satisfy the following boundary conditions;
\begin{equation}
U(x^1 + 2\pi R,x^2) = z_1P^{-m} U(x^1,x^2) P^{m}
\end{equation}
and
\begin{equation}
U(x^1,x^2 + 2\pi R) = z_2QU(x^1,x^2)Q^{-1} 
\end{equation}
where $z_1$ and $z_2$ are $Z_N$ phases $\exp{2\pi i k_1/N}$
and $\exp{2\pi i k_2/N}$ .  
The eigenvalues of these operators are 
$\exp{2 \pi i {\vec e} \cdot {\vec k}/N}$
where ${\vec e}$ are the 't Hooft electric fluxes.
An example of a large gauge transformation is
\begin{equation} U(x) = P 
\end{equation}
for which $k_1 = 0$ and $k_2 = 1$.
Under $A \rightarrow PAP^{-1}$,  the Fourier modes transform
\begin{equation}
a_{\vec r} \rightarrow a_{\vec r} e^{-2\pi i \frac{c}{N}r_1}
\label{lgt}
\end{equation}
In the dual non-commutative theory this corresponds to the translation
\begin{equation}
{x^{\prime}}^1 \rightarrow {x^{\prime}}^1+ 2\pi R^{\prime}\frac{c}{N}.
\end{equation}
Since translations in the non-commutative theory are generated by 
the momentum operator $n^{\prime}_1$, and the large gauge 
transformation (\ref{lgt}) in the commutative theory by the 
electric flux operator $\tilde e_2$,
we find
\begin{equation}
e^{2\pi i \frac{\tilde e_2}{N}} = e^{2 \pi i c \frac{n^{\prime}_1}{N}},
\label{reln}
\end{equation}
where $n^{\prime}$ is the integer quantized momentum on the non-commutative 
torus, and $\tilde e$ is the $SU(N)$ 't Hooft electric flux.
The 't Hooft flux is only defined modulo $N$.
Note that in the $U(N)$ theory,  the Wilson loop in the fundamental representation
which wraps a cycle of the torus $e_2$ times creates a $U(1)$ electric flux $e_2$
and an $SU(N)$ electric flux $\tilde e_2$ which satisfies 
$\tilde e_2 =  e_2 \ mod \ N$.  

As an aside,  one consequence of the relation (\ref{reln})
is that translation invariance in the non-commutative theory may be
spontaneously broken upon shrinking a cycle of the torus.  In the dual commutative
theory,  this corresponds to spontaneous breaking 
of a $Z_N$ symmetry upon shrinking a cycle of the dual torus.  
This phenomenon corresponds
to a finite temperature deconfinement 
transition,  which is known to occur in the pure Yang-Mills theory \cite{YS}.   

The explicit map between commutative Wilson loops and non-commutative open
Wilson lines lines with winding number and transverse 
momentum is described in detail in the appendix. We can summarize the 
map as follows.
Let us define the operator $\hat O_{n_1,e_2}$ to be the closed Wilson line
in the fundamental representation in the commutative $U(N)$ theory,  with
a transverse momentum $n_1$ and winding number  $e_2$:
\be
\hat O_{n_1,e_2} = \int dx^1 \exp(i\frac{n_1x^1}{R}) 
Tr W(x^1,x^2\rightarrow x^2 + 2\pi R e_2) 
Q^{-e_2}\exp(-i\frac{me_2 x^1}{NR}).
\label{wilcom}
\ee
Here $W$ is the path ordered exponential between the indicated points,
where the path is taken to be a straight line. 
The term ${U_2}^{-e_2}=Q^{-e_2}\exp(-i\frac{me_2 x^1}{NR})$ is the transition function required
for gauge invariance. 
Let us also define the operator $\hat O^{\prime}_{n_1^{\prime}, e_2^{\prime}}$
to be an open Wilson line of the dual non-commutative theory;
\be
\hat O^{\prime}_{n_1^{\prime}, e_2^{\prime}} = \int dx^1dx^2
W(x^1,x^2 \rightarrow x^2 + 2\pi R^{\prime} e_2^{\prime} + 
2\pi R \frac{c}{N} n_1^{\prime} ) \ast \exp(i\frac{n_1^{\prime}x^1}{R^{\prime}})\ee 
The map is then
\be
O^{\prime}_{n_1^{\prime}, e_2^{\prime}} = 2\pi RN \hat O_{n_1,e_2}
\label{themap}
\ee
where the electric charges and momenta are related by an $SL(2,Z)$ matrix;
\begin{equation}
\pmatrix{
e_2\cr
n_1
}
=
\pmatrix{
N & c \cr
m & a
}
\pmatrix{
e^{\prime}_2\cr
n^{\prime}_1
}
\label{sltz}
\end{equation}
Note that for the maximally supersymmetric case, the relation (\ref{sltz}) could have 
been arrived at from T-duality considerations in string theory.  However
we wish to emphasize that the Morita equivalence and the mapping of Wilson lines 
discussed above holds even for pure Yang-Mills theory.  

\subsection{Other Operators}

One can extend this map to include the insertion of local gauge covariant operators.  
Consider for instance the operator of the non-commutative theory given by
\be
\int d^2x {\cal F}_{\mu\nu}(x^1, x^2) *  
W(x^1,x^2 \rightarrow x^2+2\pi R^{\prime} \frac{c}{N}n_1^{\prime} 
+ 2\pi R^{\prime}e_2^{\prime}) \ast \exp(i\frac{n_1x^1}{R^{\prime}})
\ee
Here the end point of the open Wilson line begins at the location of the local operator 
insertion.  However for the 
straight Wilson line (see \cite{GHI}) the 
operator is unchanged if the local operator is inserted anywhere
within the $P^*$ ordered Wilson line.  
The commutative dual of this operator is  
\be
\int dx^1dx^2 Tr F_{\mu\nu}(x^1,x^2)W(x^1,x^2 \rightarrow x^2+2\pi R e_2)
\exp(i\frac{n_1x^1}{R}).
\label{comopins}
\ee
Note that on the commutative side, we could have integrated over $x^2$ with a 
factor $\exp( i n_2x^2/R)$. The non-commutative dual would then be an 
open Wilson lines with a non-zero longitudinal momentum. On a torus 
with rational 
$\Theta$,  it is possible to give a longitudinal momentum to the open Wilson line  
because gauge invariance only requires
the momentum factor to be equivalent to a translation operator relating the endpoints
of the open Wilson line (see figure 1). 
In addition to the longitudinal translation generated by the
transverse momentum factor, one can translate all the way around the torus in the 
transverse direction.  This corresponds to the inclusion of a longitudinal momentum
$n_2^{\prime} = Nn_2$.  

\begin{figure}
\label{translation}
\begin{center}
\epsfig{file=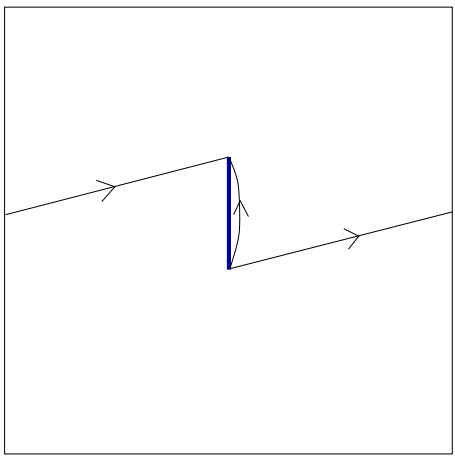,height=6cm,width=6cm}
\end{center}
\caption{ Open Wilson lines on a non-commutative 
torus admit longitudinal momenta,  
corresponding to translations all the way around the torus in
a direction transverse to the Wilson line. 
The combination of longitudinal and transverse momenta must 
correspond to a translation relating the endpoints of the 
open Wilson line.
}
\end{figure}

Similarly, we can construct the non-commutative 
duals of the $Tr F^2(x)$ operators in the commutative theory using the 
explicit map between the gauge fields in the two theories
\footnote{We thank Wati Taylor for pointing out this possibility to us.}.
A straightforward calculation gives the following operator in terms of the
Fourier modes of the field strength:
\begin{eqnarray}
Tr F^2 (x) = N
\sum_{k,n} f^{\mu \nu}_{N k-n} f^{\mu \nu}_n e^{-\pi i c N k_1 k_2}
 e^{-\pi i c (n_1 k_2-k_1 n_2)} e^{-\frac{i {\vec k}{ \vec x}}{R}}.
\end{eqnarray}
In the non-commutative theory this yields a family of gauge invariant 
operators that depend explicitly on $N$. 
There is an obvious generalization to operators $Tr F^n(x)$, which  yield
new gauge invariant operators in non-commutative gauge theories with rational
non-commutativity parameter.

\section{Continuity in $\Theta$}
\label{correlators}

Continuity in $\Theta$ depends
on the existence of very non-trivial relations between large and small
$N$ gauge theories.  For instance,  if $\Theta = -1/2$,  the commutative dual
has $N = 2$ and $m = 1$,  however if $\Theta = -1000/2001$ the commutative dual has 
$N=2001$ and $m = 2$.  
Furthermore even at the classical level, the non-commutative theory on a 
torus does 
not have a manifestly smooth dependence on $\Theta$,  since the gauge 
potential is periodic,  with periodicity given by (\ref{period}).
Nevertheless,   when considering gauge 
invariant observables we will find smooth dependence on $\Theta$,  at least 
in the instances which we have thus far considered. 

\subsection{Energies}

A first fairly trivial observation is that the BPS mass formulae in Yang-Mills 
theory \cite{CDS, Hofman, KS, PS} with
$16$ supercharges is consistent with continuity in $\Theta$.
One can ask whether the same holds true for the energy of an electric flux line
in a confining Yang-Mills theory.
Consider the  pure four-dimensional
commutative Yang-Mills theory on the space $T^2 \times S^1$ where the 
$S^1$ has size $L_3$ and the $\Theta$ parameter lives
on the $T^2$ which has radii equal to $R$.  
For sufficiently large $R$,  the principle contribution to the energy of an 
electric flux on $T^2$ comes from the confined flux lines
in the $SU(N)/Z_N$ component of the theory.
The energy of an electric flux $e_i$ is given by
\begin{equation}
E(e_i) = \Lambda^2 R e_i + \cdots.
\end{equation}
where $\Lambda$ is the QCD scale and the omitted terms
are subleading in $1/\Lambda R$. 
In the dual $U(1)$ non-commutative theory,  this is the energy of a state
created by an open Wilson line.  For concreteness, consider an electric
flux $e_2$.  Continuity in $\Theta$ in the non-commutative theory implies
that the energy of the electric flux can be written in terms of 
variables in the dual $U(1)$ non-commutative theory 
which  vary smoothly or are fixed as one varies $\Theta$.
Using formula (\ref{sltz})  the energy of the commutative electric flux can
be written as 
\begin{equation} 
E(e_2) = \Lambda^2 R^{\prime} (\frac{c}{N} n_1^{\prime} + {e}_2') + \cdots. 
\end{equation}
Thus the energy is consistent with continuity in $\Theta = -c/N$ 
(at least to leading order in $1/\Lambda R$) provided 
the QCD scale $\Lambda$ can be written in terms of 
variables in the non-commutative dual which do not differ for 
arbitarily close values of $\Theta = -c/N$.  In two-dimensional 
QCD,  the string tension is $g^2N$ or ${g '}^2$,  which can indeed 
be held fixed as one varies $\Theta$.  In four dimensions,  
we can address this question in the context of a confining vacuum 
of the ${\cal N} =1^*$ SYM theory.  
This theory is obtained by deforming the ${\cal N} =4$ theory with 
a mass term,   which in  ${\cal N} =1$ language
adds $\Delta W = \frac{1}{g^2} M Tr X^2$ to the superpotential.
Morita equivalence treats the adjoint scalars 
in the same way as the gauge fields,
so that the superpotential of the non-commutative dual is deformed by a term 
$\frac{1}{ {g '}^2 } {M '} {\cal X}^2$ with $M^{\prime} = M$. 
At very high energies,  the ${\cal N} =1^*$ theory is conformal,  with
gauge coupling $g^2$ (in the commutative description),
while below the scale $M$ the coupling runs 
according to the QCD beta-function.  For small 't Hooft coupling, the
QCD scale is given by 
\be
\Lambda = Me^{-\frac{1}{g^2N}}
\ee
This can be written in terms of non-commutative variables as
\be
\Lambda = {M '}e^{-\frac{1}{ {g '}^2}},
\ee
which is fixed as one varies $\Theta$. 

At this stage we have only checked the continuity of energies in $\Theta$
to leading order in an $1/\Lambda R$ expansion.  It may be that these
results do not persist beyond leading order. 
For a very small torus,  one could investigate the  question
of continuity in $\Theta$ perturbatively,  but we shall not do so 
here.


\subsection{Wilson loop correlators in  ${\cal N} = 4$ SYM.  }

Other quantities in the non-commutative theory 
which one might expect to depend smoothly on
$\Theta$ are the correlation functions of open Wilson lines. 
We consider the correlation functions  
\bea
G_{n_1,e_2} & \equiv & <\hat O_{n_1,e_2} \hat O_{-n_1,-e_2}> \nonumber \\
G^{\prime}_{ n_1^{\prime},e_2^{\prime} } & \equiv & 
<\hat O^{\prime}_{n_1^{\prime},e_2^{\prime}}
\hat O^{\prime}_{-n_1^{\prime},-e_2^{\prime}}>
\eea
of wrapped Wilson lines in the commutative theory and open Wilson lines
in the non-commutative theory respectively. 
Via the duality map (\ref{themap}),
continuity in $\Theta$ requires
that the correlator of closed Wilson loops wrapping the torus 
in the commutative description has the following functional form
\bea
G_{n_1,e_2} = \frac{1}{(2 \pi RN)^2}
G^{\prime}_{ n_1^{\prime},e_2^{\prime} } 
= f(\Theta, {g^{\prime}}^2, R^{\prime}, n_1^{\prime}, e_2^{\prime}) 
= \nonumber \\ 
f(-\frac{c}{N}, g^2N, RN,  Nn_1-me_2, -cn_1 + ae_2) 
\eea
Using $aN - cm =1$ and dimensional analysis,  this can be rewritten as 
\be
G_{n_1,e_2} =
(e_2R)^2 {\tilde f} (-\frac{c}{N}, g^2N, Nn_1-me_2, \frac{e_2}{N})
\label{smoothreq}
\ee
for some function $\tilde f$. 
Any other functional form would lead,  via Morita equivalence,
to a non-commutative theory in which the Wilson loop correlator
differs between arbitarily close rational values of $\Theta$.

One could try to verify continuity perturbatively,  since
Morita equivalence is a classical duality.  At leading order,
the result for the correlator in the non-commutative theory
is independent of $\Theta$ (see \cite{GHI} ). Therefore,
via Morita equivalence, the commutative theory has the right
functional behavior at leading order (\ref{smoothreq}).
It would be interesting to check whether (\ref{smoothreq})
is satisfied in a loop computation, although we will not
do this here.
In what follows, we 
shall verify the behavior predicted  by continuity in $\Theta$ 
to leading order in the $1/N$ expansion of strongly coupled 
${\cal N}=4$ $U(N)$ Yang-Mills theory.

\begin{figure}
\begin{center}
\epsfig{file=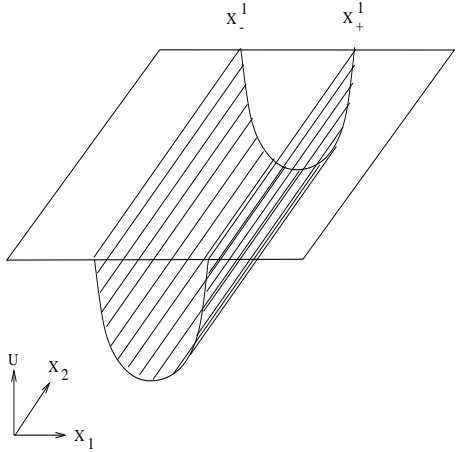,height=6cm,width=6cm}
\end{center}
\caption{ A minimal surface connecting Wilson loops on the torus, depicted
by a square with identified edges. }
\end{figure}

In the commutative ${\cal N}=4$ $U(N)$ Yang-Mills theory,  the 
correlator of Wilson loops on $T^2$ can be computed at large $N$ 
and strong 't Hooft coupling using the AdS-CFT correspondence 
\cite{adscft,WIT} and the method of \cite{Rey2} \cite{maldaloops}. 
In this case one would at least expect that when a supergravity 
dual of the non-commutative theory exists (see \cite{HI, MR}),
and  the geometry depends smoothly on $\Theta$, the formula  
(\ref{smoothreq}) ought to be satisfied. 
Note that the supergravity description is generically not
valid in all T-dual descriptions of the theory (although there may
be overlapping regions of validity).
We shall look for the smooth behavior (\ref{smoothreq}) directly in
the supergravity description of the commutative theory.  
Consider the supergravity background which is dual to 
the commutative Yang-Mills theory with 't Hooft flux $m$.
The supergravity description is valid in the large $N$ limit at fixed 
but large 't Hooft coupling.  The background is given by 
\bea
ds^2 &=& \alpha^{\prime}\left[\frac{U^2}{\sqrt{\lambda}}
(dt^2+dx_1^2 +dx_2^2 +dx_3^2) + \frac{\sqrt{\lambda} dU^2}{U^2} + 
\sqrt{\lambda}d\Omega_5^2\right] \nonumber  \\
e^{\phi} &=& \frac{\lambda}{4 \pi N} \nonumber \\
B &=& -\frac{m}{N}\frac{1}{2\pi R^2}dx^1 \wedge dx^2 
\eea
where $\lambda =4 \pi g^2 N$ and
$x^1$ and $x^2$ are compactified on a square torus of radius $R$. 
Note that the $B$ field takes precisely the value we imposed in 
section \ref{morita}
to cancel the $U(1)$ flux.
The correlator of Wilson loops can be
obtained by summing over minimal area surfaces in $AdS_5\times S_5$ 
with boundaries given by the Wilson loops at the boundary of AdS 
space. \footnote{
As in \cite{maldaloops}, we consider Wilson loop operators 
involving the scalar fields of the ${\cal N} = 4$ SYM theory: $W =
Tr \exp\left(i \int A + \epsilon_IX^I\right)$ for fixed $\epsilon$. 
This amounts 
to fixing the boundary of the minimal surface at a point in the $S^5$.
The Morita map applies as before.  Note that there is no $1/N$ in
front of the Wilson loop operator  we consider.  With this normalization
the correlator of wrapped Wilson loops is of order $1$ at large $N$.} 
Consider oppositely oriented Wilson loops wrapping the $2$-direction
$e_2$ times at the points $x^1 = x^1_+$ and $ x^1_-$  as in figure 2.

The minimal area surfaces have the form
\bea
x^2(\sigma, \tau) &=& 2\pi R e_2 \sigma \nonumber \\
x^1(\sigma, \tau) &=& (x^1_+ - x^1_- +2\pi Rn)\tau \nonumber \\
U(\sigma, \tau) &=& U(\tau),
\label{minsurf}
\eea
where $\sigma$ and $\tau$ are defined on the interval $[0,1]$ and $n$ 
is an integer characterizing the winding number of the 
minimal surface in the $x^1$ direction.  $U(\tau)$ is determined by
extremizing the world sheet action 
\be
S_n= \int \sqrt{det G} = 2\pi R e_2 
\int d\tau\sqrt{
\left(\frac{\partial U}{\partial \tau}\right)^2
+ \frac{U^4}{\lambda}
\left| x^1_+ - x^1_- + 2\pi Rn \right| ^2}.
\ee
Although the context is somewhat different,  this is precisely the same 
minimization problem as the one solved in \cite{maldaloops}.  In the limit in which
the boundaries are at $U\rightarrow \infty$,  the world sheet action
is divergent until making an appropriate subtraction.  
The result obtained after this subtraction can be extracted from 
\cite{maldaloops} and is given by
\be
S_n(x^1_+ - x^1_-) = -\frac{ 2\pi R e_2 (4\pi)^2(2\lambda)^{1/2} }
	 { \Gamma(\frac{1}{4})^4
		| x^1_+ - x^1_- + 2\pi Rn | }
\ee
In position space,  the correlator of Wilson loops
at strong 't Hooft coupling and leading order in the 
$1/N$ expansion is given formally by 
\be
G_{e_2}(x^1_+, x^1_-) =  e_2 \sum_n  
\exp\left(- i\frac{me_2(x^1_+ - x^1_- + 2\pi Rn)}{NR}\right)
\exp\left(- S_n ( x^1_+ - x^1_-) \right)
\label{woutm}
\ee

Several comments are in order. First,  the phase factor proportional
to the magnetic flux arises because of the relation 
between the flux and the NS-NS B field modulus.  Note that
for $m = 0 \ mod \ N$,  the sum on the index $n$ is divergent,
due to the long range correlations of the ${\cal N} =4$ theory.  
We shall just consider the formal sum for the moment.

Second, the overall factor of $e_2$ in
(\ref{woutm}) requires
some explanation.  The Wilson loops forming the boundary of the 
minimal surface correspond to an element of the permutation group 
${\cal S}_{e_2}$  which is a cycle of length $e_2$.  The easiest way to
understand the overall factor of $e_2$ in (\ref{woutm}) is to
infinitesimally deform both Wilson loops,  so that the $e_2$ cycles of each 
do not lie on top of each other.  For each Wilson loop, let us
label the points at $x^2 =0$ by an index $i \in\{ 1, \cdots, e_2 \}$,
 and
choose a point $i=1$ on one of the Wilson loops $\gamma^-$.  
As one follows the $x^2 =0$ path on the world sheet from $\gamma^-$ to
the other  Wilson loop $\gamma^+$,  there are $e_2$ possible end points
on $\gamma^+$.  These are topologically distinct  when
the boundaries are held fixed.
One can also understand the 
factor of $e_2$ in terms of topologically distinct planar
diagrams, as illustrated in figure 3.  This factor of $e_2$ will prove
essential for smooth dependence of the non-commutative dual on $\Theta$.   

\begin{figure}
\begin{center}
\epsfig{file=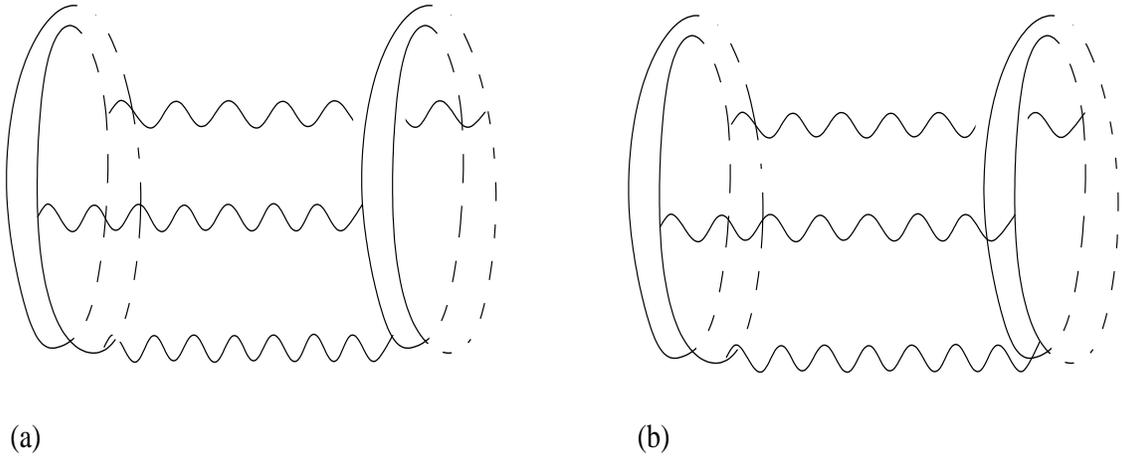,height=6cm,width=15cm}
\end{center}
\caption{ Two planar (ladder) diagrams giving an identical contribution
to correlators of Wilson loops with $e_2 =2$.}
\end{figure}

Note that due to the sum on $n$ in (\ref{woutm}),  the correlator
is periodic in $x^1_+$ and $x^1_-$.  Recall that the $SU(N)$ 
Wilson loop defined by 
(\ref{comwils}) is not periodic in the presence of 't Hooft flux \cite{VB}, 
whereas the $U(N)$ Wilson loop is periodic. In the commutative case, 
the AdS-CFT relation  is
said to describe an $SU(N)/Z_N$ rather than $U(N)$ theory (e.g. \cite{WIT}).
 We therefore interpret the result (\ref{woutm}) as the correlator of 
$U(N)$ Wilson loops with a non-dynamical background $U(1)$,  
with first Chern class
equal to the $SU(N)$ 't Hooft flux.  
We now Fourier transform (\ref{woutm}), giving the correlator of Wilson 
loops with momentum $n_1$:
\bea
G_{n_1,e_2} &=& \int_0^{2\pi R} dx^1_+ \int_0^{2\pi R}dx^1_-  
e^{i\frac{n_1(x^1_+ - x^1_-)y}{R}} G_{e_2}(x^1_+, x^1_-)
 \nonumber \\
&=& 2\pi R e_2 \int_0^{2\pi R} dy
\exp (i\frac{n_1y}{R}) \label{limt} \\
& & \sum_{n=-\infty}^{+\infty} 
\exp\left(-i\frac{me_2(y + 2\pi Rn)}{NR} \right)
\exp\left(\frac{ 2\pi R e_2 (4\pi)^2(2\lambda)^{1/2} }
	 { \Gamma(\frac{1}{4})^4
		| y + 2\pi Rn | } \right)  \nonumber \\
&=& 2\pi Re_2 \int_{-\infty}^{+\infty}dy 
\exp\left(i\frac{(n_1N - me_2) y }{NR} \right) 
\exp\left(\frac{ 2\pi R e_2 (4\pi)^2(2\lambda)^{1/2} }
	 { \Gamma(\frac{1}{4})^4
		| y | } \right)
\nonumber
\eea
Note that this converges as long as the momentum 
$n_1N -me_2$ is non-zero.  The zero momentum divergence
arises because  the ${\cal N} =4$
theory is conformal.  The correlator (\ref{limt}) has the functional form
\be
G_{n_1,e_2} = (e_2 R)^2 f(g^2N, (n_1N-me_2)\frac{e_2}{N})
\ee
which is consistent with continuity in $\Theta$. 
It would be very interesting to explicitly see if the $\frac{1}{N}$
corrections are also consistent with continuity in $\Theta$.    
Very little is known about the $1/N$ corrections in the 
strongly coupled ${\cal N} = 4$ super Yang-Mills theory.
Still, the behavior of the leading term in the large 
$N$ result is by itself non-trivial.  

Note that one expects the Wilson loop correlator in the commutative
theory to depend smoothly on $m/N$ in
't Hooft large $N$ limit at fixed $e_2$, $n_1$ and $R$.
This excludes explicit dependence on 
$-c/N$ in the $1/N$ expansion.  Assuming this and continuity
in  $\Theta$ implies a $1/N$  expansion of the form
\be
G_{n_1,e_2} = (Re_2)^2 \sum_l \left(\frac{e_2}{N}\right)^l
g_l\left(g^2N, \frac{e_2}{N} (Nn_1 - me_2)\right)
\label{Nexpansion}
\ee  
In writing this expression we have also made use of the fact that
$Nn_1 -me_2$ grows like $N$ for a generic choice of $n_1$ and $e_2$,  
while the argument of $g_l$ is by definition  of order one.

The above discussion has some interesting consequences for
gauge theory on a non-commutative $R^2$ in the 
$\theta \rightarrow \infty$ limit.
It has been pointed out that the $\theta \rightarrow \infty$ limit
of a field theory on a non-commutative $R^2$ behaves like
a string theory, due to the vanishing of all but the planar graphs
\cite{MRS}.
In the $N \rightarrow \infty$ limit of the commutative theory 
with $R, e_2, n_1$ and $m/N$ fixed,  the 
correlator (\ref{limt}) can be interpreted as that of an
open Wilson line with fixed length and transverse momentum in
the strong coupling limit of the $\theta \rightarrow \infty$ 
theory,  where $\theta = \Theta 2\pi {R^{\prime}}^2$ and 
$R^{\prime} = RN \rightarrow \infty$.
The length of the open Wilson line is 
\be
L^{\prime} = \Theta n_1^{\prime} + 2\pi R^{\prime}e_2^{\prime} = e_2R
\label{length}
\ee
while the momentum is
\be
p^{\prime} = \frac{n_1^{\prime}}{R^{\prime}} = 
\frac{(Nn_1 -me_2)}{NR},
\ee  
both of which are fixed in the large $N$ limit.
It may seem odd that open Wilson lines of arbitrary
momentum $p$ and length $L$ 
exist in the infinite $\theta$ and infinite $R^{\prime}$ limit 
of the non-commutative theory,  where one would  
naively write $L^{\prime} = \theta p^{\prime}$.
However the non-commutative electric flux $e_2^{\prime}$ 
can conspire to give a finite length in this limit,
since the length is actually given by (\ref{length}).

\subsection{Wilson loop correlators  in confining Yang-Mills}

The Morita equivalence (\ref{themap}) applies
classically even in the pure $U(N)$ Yang-Mills theory. 
This raises the question of whether confining Yang-Mills theories 
exhibit behavior consistent with continuity in $\Theta$.
These theories differ greatly from the ${\cal N} =4$ theory,  
and in the non-commutative case exhibit $UV/IR$ mixing \cite{MRS}.
Furthermore, it is difficult to embed asymptotically free confining 
theories in string theory.  
It may be that the ${\cal N} =4 $ case exhibits continuity in
$\Theta$ while the confining theories do not.  
We shall address this question of continuity in $\Theta$ 
in the $1/N$ expansion of confining 
Yang-Mills and find evidence for continuity at leading order.   
  
Based in large part on observations of 't Hooft \cite{planar}, 
the large $N$ limit of QCD is expected to be described by a string theory.  
In two dimensions,  where an exact 
solution exists \cite{Rusakov},  the string description has been
constructed \cite{KK, Gr, GT1, GT2, CMR, Hor}.
Let us consider this situation first.  We will again consider
the correlation function of Wilson loops $\hat O_{n_1,e_2}$ 
wrapping the torus.  
The correlator can be computed using methods described in \cite{GT2, CMR}.  
The basic idea is that one sums over maps without folds from the 
world sheet to the target space, $\Sigma \rightarrow T$, such
that the boundary $\partial \Sigma$ is a cyclic $e_2$-fold cover of
the $S^1$ associated to each Wilson loop.  At leading order in the
$1/N$ expansion,  one sums over covers without branch points.
The 't Hooft magnetic flux plays the role of
a $B$ field modulus of the 2-d QCD string. 
This means that in the sum over
maps of the worldsheet to the target space $\Sigma \rightarrow T$,
string worldsheets are weighted by a phase factor
\be
e^{-i \frac{m}{N}\frac{1}{2\pi R^2} \int_\Sigma dx^1 \wedge dx^2} 
\ee
just as in the AdS-CFT case. In 2-d QCD this can be seen for instance
by computing with the heat kernel action  with  twisted 
boundary conditions.
The correlation function at leading order in $1/N$ is given by
\bea
G_{n_1,e_2} &=&
2\pi R e_2 \int_0^{2\pi R} dy
\exp(i\frac{n_1y}{R}) \nonumber \\
& & \sum_{n=-\infty}^{+ \infty} 
\exp\left(-i\frac{me_2(y + 2\pi Rn)}{NR} \right)
\exp\left(-\lambda 2\pi R e_2 |y+ 2\pi Rn| \right)
\label{confcor}
\eea
where $\lambda$ is the 't Hooft coupling $g^2N$ and 
$n$ indicates a winding number of the world sheet in the direction
transverse to the Wilson loops.  The overall factor of $e_2$ arises
for the same reasons as in the AdS-CFT case.
The above expression can be rewritten as
\bea
G_{n_1,e_2} =
2\pi R e_2 \int_{-\infty}^{\infty} dy
\exp(i\frac{(n_1N-me_2)y}{NR})
\exp\left(-\lambda 2\pi R e_2 |y| \right)
\label{conf}
\eea
This is manifestly consistent with smooth dependence on
$\Theta$, as it depends only on the variables $e_2R, g^2N,
NR$ and $n_1N-me_2$\footnote{While the Wilson loop correlation 
function is smooth,  the partition function (or vacuum energy) 
is not, since it
it depends on the variable $m/N + i\lambda R^2$ 
\cite{zt},  which
can not be written in terms of non-commutative variables which 
vary continuously with $\Theta$.}.  
At this point we have been somewhat cavalier about the distinction 
between $U(N)$ and $SU(N)$.
The above correlator is that of an $SU(N)$ Wilson loop,  with
an extra $U(1)$ phase factor in the transition function such
that the Wilson loop is periodic in the transverse direction.
This can be interpreted as a $U(N)$ correlator with 
a frozen $U(1)$,  much like the correlator in the AdS-CFT case.
At higher orders in the $1/N$ expansion, the distinction between 
$U(N)$ and $SU(N)$ will certainly be important.  
The role of the $U(1)$ is at this stage somewhat mysterious. 
It would be interesting to compute the $1/N$ corrections, and
see if they are consistent with continuity in $\Theta$. The $1/N$
corrections arise by considering branched covers,  as well as 
collapsed handles and tubes in the $SU(N)$ case.

One can also make similar statements about continuity in $\Theta$
in four dimensional confining Yang-Mills theories. The argument
depends crucially on the relation of the 't Hooft magnetic flux
to a B field modulus of the 4-d QCD string.  This relation can be shown
\cite{zflux} in the context of MQCD \cite{MQCD} and its IIA limit 
where it arises as a consequence of a freezing of the $U(1)$ degree
of freedom.  In the MQCD context,  the correlator can be computed by
summing over M2-branes with boundaries on the Wilson loop 
times an interval (see \cite{zflux} ).  At leading order the result
is identical to (\ref{conf}) with the replacement 
$\lambda \rightarrow \Lambda^2$, where $\Lambda^2$ is the MQCD string
tension.

\section{Conclusions and discussion}
\label{cd}

The existence of renormalizable gauge theories
on non-commutative spaces is  still an open
question.  For rational $\Theta$,  the theory may be defined
on the torus via Morita equivalence with gauge theory on
a commutative torus.  However it is not manifest that
observable quantities are arbitrarily close
for arbitrarily close values of $\Theta$,  as one would
require to define the theory for irrational $\Theta$.
For this to be the case,  there must be strong constraints
on observables in commutative Yang-Mills theories.
In other words, if the non-commutative gauge theory on the
torus could be defined for all values of $\Theta$,
consistent with Morita equivalence, we would have strong predictions for
commutative gauge theories on tori.

The strong constraints on commutative gauge theories
 are easily seen to be satisfied for such quantities
as the energies of BPS states in Yang-Mills theories with $16$
supercharges.
In this paper we have verified that
these constraints are also obeyed by the correlator of
wrapped Wilson loops, to leading order in the $1/N$
expansion  of four dimensional
${\cal N} =4$ Yang-Mills theory,  and pure
two dimensional QCD.
We  also presented rough arguments
concerning the behavior of this correlator in confining
four dimensional QCD,  where it again seems to be consistent
with continuity in $\Theta$.

While the leading $1/N$ behavior of the correlator is consistent
with continuity in $\Theta$,  there exist other quantities which
do not vary continuously with $\Theta$.  Amongst these is
 the periodicity of the
non-commutative gauge potential, and certain observables which can
only be defined for rational values of $\Theta$.
Moreover, in the two-dimensional case, continuity in
$\Theta$ is only exhibited by the correlator,  but not
by the partition function.

To check further whether commutative gauge theories satisfy the
constraints implied by smooth non-commutative duals,
it would be  enlightening to study the subleading terms in
the $1/N$ expansion in cases in which this is possible,
such as two-dimensional QCD. Perhaps a next to leading order term
can shed light on whether the pattern observed in this paper persists.
If it persists, it yields strong predictions on commutative gauge
theories. If it does not, there is a puzzle as to the definition of
non-commutative gauge theories on the torus for irrational values of
$\Theta$.  It would also be interesting to check continuity in 
$\Theta$ perturbatively.  In the confining case,  this can be 
done on a sufficiently small torus.

A further point of investigation could be the following.
Using the AdS/CFT correspondence, we computed the correlator of closed
Wilson loops in the large $N$ limit. We made good use of the fact that
the correlator of spatially separated Wilson loops is equivalent to a
computation of the quark-anti-quark potential with a compactified
time direction (where
the compactification is assumed to preserve
supersymmetry). That allowed us to recuperate the results in
\cite{maldaloops}.
We expect this correlator to be equivalent to a correlator of open Wilson
lines in a non-commutative $U(p)$ theory
that we should be able to compute at large $p$ using the non-commutative
version of the AdS/CFT correspondence \cite{Jabbar, DK}.
 Since these correlators are dual,
they
should yield the same result, after dualising momentum and winding number
of the loops. 
 It should be instructive to try and follow
 the AdS/CFT
correspondence
(for theories compactified on tori) under the Morita equivalence map, to
learn more about the correspondence for non-commutative field
theories.

\section*{Acknowledgements}

We are grateful to Allan Adams, Ori Ganor, Ami Hanany,  Aki Hashimoto,  
Roman Jackiw,  Anatoly Konechny, Sanjaye Ramgoolam, Wati Taylor, 
Pierre van Baal and 
Frank Wilcek  for helpful conversations. This work is supported in part
by funds provided by the U.S. Department of Energy (D.O.E.) under
cooperative research agreement DE-FC02-94ER40818.


\section{Appendix: mapping Wilson lines}

In this appendix we prove the relation (\ref{themap}) discussed in
\cite{AMNS, S},  taking into account Wilson lines with both winding
and transverse momentum.
We will do so by expanding the path ordered exponentials in
$\hat O^{\prime}_{n_1',e_2'}$ and $2\pi RN \hat O_{n_1,e_2}$ as follows:
\bea
 P(e^{i\int A}) = \ \ \ \ \ \ \ \ \ \ \ \ \ \ \ \ \ \ \ \   
\ \ \ \ \ \ \ \ \ \ \ \ \ \ \ \ \ 
\ \ \ \ \  \ \ \ \ \ \ \ \ \ \ \ \ \ \ \ \ \ \ 
\ \ \ \ \ \ \ \ \ \ \ \ \ \ \ \ \ \ \ \ \ \ \ \ \ \ \ \nonumber \\  
1 + i\int_0^1 ds A_{\mu}(x(s))\frac{dx^{\mu}(s)}{ds}
-\int_0^1 ds \int_0^s d {\tilde s}A_{\mu}(x(s))A_{\nu}(x({\tilde s}))
\frac{dx^{\mu}(s)}{ds}\frac{dx^{\nu}(\tilde s)}{d \tilde s} +\cdots 
\nonumber \\
\ \ \ 
\label{expath}
\eea
and showing equivalence at 
all orders.  
Inserting this expansion into the expression (\ref{wilcom}) for
the operator $2\pi RN \hat O$, and using expression (\ref{genform}) for $A$,
gives 
\bea
\int_0^{2 \pi R} dx^1 \int_0^1 ds_a \int_0^{s_a} ds_b \cdots  
\exp \left(
-2\pi i 
( r_{2a}s_a +r_{2b}s_b + \cdots)
\frac{e_2}{N} 
\right) 
\nonumber \\
\exp\left(-i\pi \frac{c}{N}(r_{1a}r_{2a} + r_{1b}r_{2b} + \cdots)\right)  
Tr \left(Q^{-c r_{1a}}P^{r_{2a}}Q^{-c r_{1b}}P^{r_{2b}} 
\cdots Q^{-e_2}\right)
\nonumber \\
\exp\left(-i(\vec r_a \cdot \vec x + \vec r_b \cdot \vec x \cdots 
+ me_2x^1)/NR\right)
\exp(in_1x^1/R) a^2_{\vec r_a}a^2_{\vec r_b}\cdots \nonumber \\
2\pi R N (2\pi R e_2)^k
\label{start}
\eea
where we have used
\be
x^2(s) = x^2 + 2\pi Re_2 s
\ee
for $s$ defined on the interval $[0,1]$ and $k$ denotes the order in the
expansion of the exponential.
We wish to show that (\ref{start}) is equivalent order by order to
the analogous expansion for $\hat O$, given by
\bea
\nonumber \\
\int_0^{2\pi R^{\prime}} dx^1 \int_0^{2\pi R^{\prime}} dx^2 
\int_0^1 ds_a \int_0^{s_a} ds_b \cdots \nonumber \\ 
\exp \left(
-2\pi i 
( r_{2a}s_a +r_{2b}s_b + \cdots)
 (e_2^{\prime} + \frac{c}{N}n_1^{\prime} ) 
\right)
\nonumber \\
\exp\left(-i\frac{(\vec r_a \cdot \vec x)}{R^{\prime}} \right)
* 
\exp\left(-i\frac{(\vec r_b \cdot \vec x)}{R^{\prime}} \right)
\cdots \nonumber \\
* \exp\left( i \frac{ n_1^{\prime} x^1}{R^{\prime}} \right)
a^2_{\vec r_a}a^2_{\vec r_b}\cdots \left( 2\pi R^{\prime}
( e_2^{\prime} + \frac{c}{N}n_1^{\prime}) \right)^k
\nonumber \\
\ \ 
\label{ncstart}
\eea 

We will begin with (\ref{start})
and reorder the quantities within the trace,  giving
\bea 
Tr \left(Q^{-c r_{1a}}P^{r_{2a}}Q^{-c r_{1b}}P^{r_{2b}} 
\cdots Q^{-e_2}\right) &=& \nonumber \\
Tr\left(Q^{-c r_{1a} -c r_{1b} 
+\cdots -e_2} P^{r_{2a} + r_{2b} + \cdots}\right) 
\phi(\vec r_a,\vec r_b\cdots, me_2) & &
\label{defphi}
\eea
where $\phi$ is a $Z_N$ phase.  It will not be necessary to 
write it explicitly.     
Now one can readily verify that the quantity
\be 
Tr\left(Q^{-c r_{1a}-c r_{1b} 
+\cdots -e_2} P^{r_{2a} + r_{2b} + \cdots}\right) 
\ee
vanishes unless 
\be
-c(r_{1a} + r_{1b} +\cdots - me_2) = kN
\label{condo}
\ee
and
\be
-c(r_{2a} + r_{2b} + \cdots ) = jN
\label{condt}
\ee
for integers $k$ and $j$.
  If these conditions are satisfied then, since $P^N = Q^N =1$,
\be
Tr\left(Q^{-c r_{1a}-c r_{1b} 
+\cdots -e_2} P^{r_{2a} + r_{2b} + \cdots}\right) = N
\ee
The Wilson loop is independent of the base point $x^2$,
  so all contributions from non-zero $j$ 
vanish. 

Let us now consider the integral over the $x^1$ 
dependent parts of (\ref{start}),
\be
\int dx^1 
\exp\left(-i\frac{(r_{1a} + r_{1b} + \cdots 
+ me_2 - n_1N) x^1}{NR} \right).
\label{xoneint}
\ee
Since $gcd(N,c) =1$ via (\ref{relprime}),
we can rewrite (\ref{condo}) as
\be
r_{1a} + r_{1b} +\cdots - me_2 =  q N
\label{condoagain}
\ee
where $q$ is an integer.
Therefore (\ref{xoneint}) is an integral over a periodic 
function,  giving 
\be
2\pi R\delta_{0,r_{1a} + r_{1b} + \cdots + me_2 - n_1N}
\ee
We can therefore write (\ref{start})
as 
\bea
\int_0^{2\pi R} dx^1 \int_0^{2\pi R} dx^2 
\int_0^1 ds_a \int_0^{s_a} ds_b \cdots  
\exp \left(
-2\pi i 
( r_{2a}s_a +r_{2b}s_b + \cdots)
\frac{e_2}{N} 
\right) \nonumber \\
\exp\left(-i\pi \frac{c}{N}(r_{1a}r_{2a} + r_{1b}r_{2b} + \cdots)\right)
\exp\left(-i\frac{(r_{1a} + r_{1b} +\cdots + me_2 - n_1N) x^1}{R}
\right) \nonumber \\
\exp\left(i\frac{(r_{2a} + r_{2b} + \cdots)x^2}{R}\right)
\phi(\vec r_a,\vec r_b\cdots, me_2)
a^2_{\vec r_a}a^2_{\vec r_b}\cdots
\frac{N}{2\pi R} 2\pi R N (2\pi R e_2)^k \nonumber \\
\ \
\eea 
This expression is equivalent to the following integral over the 
non-commutative torus;
\bea
\int_0^{2\pi R^{\prime}} dx^1 \int_0^{2\pi R^{\prime}} dx^2 
\int_0^1 ds_a \int_0^{s_a} ds_b \cdots  
\exp \left(
-2\pi i 
( r_{2a}s_a +r_{2b}s_b + \cdots)
\frac{e_2}{N} 
\right) \nonumber \\
\exp\left(-i\pi \frac{c}{N}(r_{1a}r_{2a} + r_{1b}r_{2b} + \cdots)\right)
\exp\left(i\frac{(r_{1a} + r_{1b} +\cdots + me_2 - n_1N) x^1}
{R^{\prime}}
\right) * \nonumber \\
\exp\left(-i\frac{(r_{2a} + r_{2b} + \cdots)x^2}{R^{\prime}}\right) 
\nonumber \\ 
\phi(\vec r_a,\vec r_b\cdots, me_2)
a^2_{\vec r_a} a^2_{\vec r_b}\cdots (2\pi R e_2)^k \nonumber \\
\ \
\label{newstart}
\eea 
where we have used $R^{\prime} = RN$ and the fact that 
$\int d^2 x \, A*B = \int d^2 x \, A.B$.
We can now make use of a relation between 
the $\ast$-algebra and the 
matrix algebra generated
by $Q$ and $P$.  Note that
\be
Q^{-c r_1} P^{r_2} = P^{r_2} Q^{-c r_1} 
\exp(2\pi i \frac{c}{N}r_1r_2)
\ee
and 
\bea
\exp(-i\frac{r_1x^1}{R^{\prime}}) * 
\exp(-i\frac{r_2x^2}{R^{\prime}}) = \nonumber \\
\exp(-i\frac{r_2x^2}{R^{\prime}}) * \exp(-i\frac{r_1x^1}{R^{\prime}}) 
\exp(2\pi i \frac{c}{N}r_1r_2)
\eea
Thus 
\bea
\exp\left( -i \frac{r_{1a}x^1}{R^{\prime}} \right) *
exp\left( -i \frac{r_{2a}x^2}{R^{\prime}} \right) * 
exp\left( -i \frac{r_{1b}x^1}{R^{\prime}} \right) *
exp\left( -i \frac{r_{2b}x^2}{R^{\prime}} \right) *\cdots \nonumber \\
*\exp\left(-i\frac{me_2 x^1}{R^{\prime}}\right) = \nonumber \\
\exp\left(-i \frac{(r_{1a} + r_{1b} + \cdots +me_2 - n_1N)x^1}{R^{\prime}}
\right) *
\exp\left(-i \frac{(r_{1a} + r_{1b} + \cdots)x^2}{R^{\prime}}
\right) \nonumber \\
\phi(\vec r_a,\vec r_b, \cdots, me_2) \nonumber \\
\ \ \ 
\label{alg}
\eea
where,  since $Q^{-e_2} = Q^{-c m e_2}$,  the phase $\phi$ in
(\ref{alg}) is the same as that which appears in (\ref{defphi}).
Furthermore we have
\be
\exp\left( -i \frac{r_{1}x^1}{R^{\prime}} \right) *
exp\left( -i \frac{r_{2}x^2}{R^{\prime}} \right) =
\exp\left(-i\frac{\vec r \cdot \vec x}{R^{\prime}}\right)
\exp\left(i \frac{c}{N} \pi r_1r_2\right).
\ee
Therefore (\ref{newstart}) may be written as
\bea
\int_0^{2\pi R^{\prime}} dx^1 \int_0^{2\pi R^{\prime}} dx^2 
\int_0^1 ds_a \int_0^{s_a} ds_b \cdots  
\exp \left(
-2\pi i 
( r_{2a}s_a +r_{2b}s_b + \cdots)
\frac{e_2}{N} 
\right)
\nonumber \\
\exp\left(-i\frac{\vec r_a \cdot \vec x}{R^{\prime}} \right)
* 
\exp\left(-i\frac{\vec r_b \cdot \vec x}{R^{\prime}} \right)
\cdots
*
\exp\left( -i \frac{(me_2 - n_1N) x^1}{R^{\prime}} \right)
\nonumber \\ 
a^2_{\vec r_a} a^2_{\vec r_b}\cdots (2\pi R e_2)^k \nonumber \\
\label{startagain}
\ \ 
\eea 
Formula (\ref{startagain}) is now identical to (\ref{ncstart})
with the identification
\be
\pmatrix{
e_2^{\prime}\cr
n_1^{\prime}
}
=
\pmatrix{
a & -c \cr
-m & N
}
\pmatrix{
e_2\cr
n_1}
\ee
This completes the proof.

\end{document}